\documentstyle[prl,twocolumn,aps]{revtex}
\begin{document}
\input{epsf.tex}
\epsfverbosetrue

\draft

\twocolumn[\hsize\textwidth\columnwidth\hsize\csname
@twocolumnfalse\endcsname

\title{Rotating Optical Soliton Clusters}

\author{Anton S. Desyatnikov$^{1,2}$ and Yuri S. Kivshar$^1$}

\address{$^1$Nonlinear Physics Group, Research School of Physical Sciences
and Engineering, The Australian National University, Canberra ACT 0200,
Australia}

\address{$^2$Department of Optoelectronics, Faculty of Physics,
Dnipropetrovsk National University, Dnipropetrovsk 49625, Ukraine}


\maketitle

\begin{abstract}
We introduce the concept of {\em soliton clusters} --
multi-soliton bound states in a homogeneous bulk optical medium,
and reveal a key physical mechanism for their stabilization
associated with a staircase-like phase distribution that induces a
net angular momentum and leads to cluster rotation. The ringlike
soliton clusters provide a nontrivial generalization of the
concepts of two-soliton spiraling, optical vortex solitons, and
necklace-type optical beams.
\end{abstract}

]

\narrowtext

Recent progress in generating spatial optical solitons in nonlinear bulk media opens the
possibility to study truly two-dimensional self-trapping of light and interaction of
multi-dimensional solitary waves \cite{Science}. Robust nature of spatial solitons that they
display in interactions allows us to draw a formal analogy with  atomic physics and treat spatial
solitons as ``atoms of light''. Our motivation here is to find out whether more complex objects,
viewed as ``atom clusters'', can be constructed from a certain number of simple solitons --
``atoms''. In this Letter, we describe, for the first time to our knowledge, the basic principles
for constructing the so-called ``soliton clusters'', ringlike multi-soliton bound states in a bulk
media.

First, in order to discuss the formation of multi-soliton bound states in a homogeneous bulk
medium, we should recall the physics of the coherent interaction of two spatial solitons. It is
well known \cite{Science} that such an interaction depends crucially on the relative soliton
phase, say $\theta$, so that two solitons attract each other for $\theta =0$, and they repel each
other for $\theta = \pi$. For the intermediate values of the soliton phase, $0<\theta<\pi$, the
solitons undergo an energy exchange and display inelastic interaction. As a result, {\em no
stationary bound states of two coherently interacting solitons are possible in a bulk medium.}

{\em Soliton spiraling} was suggested theoretically \cite{spiral} and observed experimentally
\cite{spiral1} as a possible scenario for a {\em dynamical} two-soliton bound state formed when
two solitons are launched with {\em initially twisted trajectories}. However, it was clarified
later \cite{spiral2} that the experimental observation of the soliton spiraling was possible due
to an effectively {\em vectorial} beam interaction. As a matter of fact,  the soliton spiraling
reported in Refs. \cite{spiral1,spiral2} is associated with large-amplitude oscillations of a
dipole-mode vector state \cite{dipole} generated by the interaction of two initially mutually
incoherent optical beams.

In spite of the fact that no bound states exist for two coherently
interacting scalar solitons in a bulk medium, in this Letter we
demonstrate that such bound states (or {\em ringlike clusters})
are indeed possible for larger number of solitons, namely for
$N\geq 4$. The main reason for the existence of such multi-soliton
states can be explained with the help of simple physics. Indeed,
let us analyze possible {\em stationary configurations} of $N$
coherently interacting solitons in (2+1)-dimensions. The only
finite-energy structures that would balance out the
phase-sensitive coherent interaction of the neighboring solitons
should possess a ringlike geometry. However, a ringlike
configuration of $N$ solitons will be {\em radially unstable} due
to an effective tension induced by bending of the soliton array.
Thus, a ring of $N$ solitons will collapse, if the mutual
interaction between the neighboring solitons is attractive, or
otherwise expand, resembling the expansion of the necklace beams
\cite{necklace}. Nevertheless, a simple physical mechanism will
provide stabilization of the ringlike configuration of $N$
solitons, if we introduce {\em an additional phase} on the scalar
field that twists by $2\pi m$ along the soliton ring. This phase
introduces {\em an effective centrifugal force} that can balance
out the tension effect and stabilize the ringlike soliton cluster.
Due to a net angular momentum induced by such a phase
distribution, such soliton clusters rotate with an angular
velocity which depends on the number of solitons and phase charge
$m$.

To build up the theory of the soliton clusters, we consider a coherent superposition of $N$
solitons with the envelopes $G_{n}(x,y,z)$, $n=1,2..N$, propagating in a self-focusing homogeneous
bulk medium. The equation for the slowly varying field envelope $E = \sum G_{n}$ can be written in
the form of the nonlinear Schr\"odinger equation,
\begin{equation}
\label{eq1}\
i\frac{\partial E}{\partial z}+\Delta _{\perp }E+f(I)E=0,
\end{equation}
where $\Delta _{\perp}$ is the transverse Laplacian and $z$ is the
propagation distance measured in the units of the diffraction
length. Function $f(I)$ describes the nonlinear properties of an
optical medium, and it is assumed to depend on the total beam
intensity, $I=\left|E\right| ^{2}$.

General features of the dynamical system (\ref{eq1}) are
determined by its conservation laws, or {\em integrals of motion}:
the {\em beam power}, $P=\int\left|E\right|^{2}d{\bf r}$, the {\em
linear momentum}, ${\bf L}={\rm Im} \int E^{\ast}\nabla E d{\bf
r}$, and the {\em angular momentum}, $M{\bf e}_{z}={\rm Im}\int
E^{\ast} ({\bf r} \times \nabla E) d{\bf r}$. For a ring of
identical {\em weakly overlapping} solitons launched in parallel,
we can calculate the integrals of motion employing a Gaussian
ansatz for a single beam $G_{n}$,
\begin{equation}
\label{ans}\
G_{n}=A\exp\left(-\;\frac{|{\bf r} -{\bf r}_{n}|^{2}} {2a^{2}}+
i\alpha_{n}\right),
\end{equation}
where ${\bf r}_{n}=\lbrace x_{n}; y_{n}\rbrace$ defines the position of the soliton center, and
$\alpha_{n}$ is the phase of the $n$-th beam. Then, the integrals of motion take the form:
\begin{equation}
\begin{array}{l}{\displaystyle
P=\pi a^{2}A^{2} \sum_{n,k=1}^N e^{Y_{nk}}\cos\theta_{nk},}\\*[9pt] \label{linmom}\ {\displaystyle
{\bf L} = \frac{\pi}{2} A^{2} \sum_{n,k=1}^N e^{Y_{nk}}({\bf r}_{n} - {\bf
r}_{k})\sin\theta_{nk},}\\*[9pt] {\displaystyle M = \pi A^{2} \sum_{n,k=1}^N e^{Y_{nk}}\vert{\bf
r}_{n}\times{\bf r}_{k}\vert\sin\theta_{nk},}
\end{array}
\end{equation}
where $Y_{nk}=-|{\bf r}_{n} - {\bf r}_{k}|^{2}/4a^2$ and $\theta_{nk}=\alpha_{n}-\alpha_{k}$. We
assume that the beams are arranged in {\em a ring-shaped array}: ${\bf r}_{n}=\lbrace
{R}\cos\varphi_{n}; {R}\sin\varphi_{n} \rbrace$ with $\varphi_{n}=2\pi n/N$, for which we find
$Y_{nk}=-({R}/a)^{2}\sin^{2}[\pi(n-k)/N]$.

First of all, analyzing many-soliton clusters, we remove the centre-of-the mass motion and take
${\bf L}=0$. Applying this constraint to Eqs. (\ref{linmom}), we find the conditions for  the
soliton phases, $\alpha_{i+n}-\alpha_{i}=\alpha_{k+n}-\alpha_{k}$, which are easy to satisfy
provided the phase $\alpha_{n}$ has a linear dependence on $n$, i.e. $\alpha_{n}=\theta n$, where
$\theta$ is the relative phase between two neighboring solitons in the ring. Then, we employ the
phase periodicity condition taken in the form $\alpha_{n+N}=\alpha_{n}+2\pi m$, and find:

\vspace{-2mm}

\begin{equation}
\label{phase}\
\theta=\frac {2\pi m}{N}.
\end{equation}

In terms of the classical fields, Eq. (\ref{phase}) gives the condition of the vanishing energy
flow ${\bf L} = 0$, because the linear momentum ${\bf L}=\int {\bf j} d{\bf r}$ can be presented
through the local current ${\bf j} = {\rm Im} ( E^{\ast}\nabla E)$. Therefore, Eq. (\ref{phase})
determines a nontrivial phase distribution for the effectively elastic soliton interaction in the
ring. In particular, for the well-known case of two solitons ($N=2$), this condition gives only
two states with the zero energy exchange, when $m$ is even ($\theta = 0$, mutual attraction) and
when $m$ is odd ($\theta =\pi$, mutual repulsion) \cite{Science}.

For a given $N>2$, the condition (\ref{phase}) predicts the
existence of a discrete set of allowed stationary states
corresponding to the values $\theta = \theta ^{(m)}$ with $m=0,\pm
1, \ldots, \pm (N-1)$. Here two states $\theta ^{(\pm |m|)}$
differ only by the sign (direction) of the angular momentum,
similar to the case of vortex solitons. Moreover, for any positive
(negative) $m_{+}$ within the domain $\pi <|\theta| < 2\pi$,  one
can find the corresponding negative (positive) value $m_{-}$
within the domain $0<|\theta|<\pi$, so that both $m_{+}$ and
$m_{-}$ describe the same cluster. For example, in the case $N=3$,
three states with zero energy exchange are possible:
$\theta^{(0)}=0$, $\theta^{(1)}=2\pi/3$, and
$\theta^{(2)}=4\pi/3$, and the correspondence is $\theta ^{(\pm
1)} \leftrightarrow \theta ^{(\mp 2)}$.  Therefore, it is useful
to introduce the main value of $\theta$ in the domain $0 \leq
\theta \leq \pi$, keeping in mind that all allowed states inside
the domain $0< \theta < \pi$ are degenerated with respect to the
sign of the angular momentum. The absolute value of the angular
momentum vanishes at both ends of this domain, when $m=0$, for any
$N$, and when $m=N/2$, for even $N$. The number $m$ determines the
full phase twist around the ring, and it plays a role of the
topological charge of a phase dislocation associated with the
ring.

In order to demonstrate the basic properties of the soliton
clusters for a particular example, we select the well-known
saturable nonlinear Kerr medium with $F(I)=I(1+sI)^{-1}$, where
$s$ is a saturation parameter. This model supports stable
(2+1)-dimensional solitons.  First, we apply the variational
technique to find the parameters of a single soliton described by
the ansatz (\ref{ans}), and find $A=3.604$ and $a=1.623$ for
$s=0.5$. Then, substituting Eq. (\ref{ans}) into the system
Hamiltonian, we calculate the effective interaction potential
$U(R)=H(R)/|H(\infty)|$, where $H$ is the system Hamiltonian,
\[
H = \int \left\{|\nabla E|^2
- \frac{1}{s}|E|^2+ \frac{1}{s^2}\ln\left(1+s|E|^2\right)\right\} d{\bf r}.
\]
As a result, for any $N$ we find {\em three distinct types} of the
interaction potential $U(R)$, shown in Fig. 1 for the particular
case $N=5$. Only one of them has a local minimum at finite $R$
which indicates the cluster stabilization against collapse or
expansion.

\begin{figure} \setlength{\epsfxsize}{7 cm}
\centerline{ \epsfbox{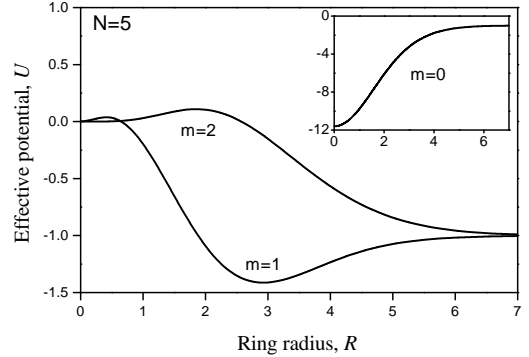}} \vspace{1.5 mm}
\caption{Examples of the effective potential $U(R)$ for a circular
array of $N=5$ solitons. Corresponding values of the topological
charge $m$ are shown near the curves. Dynamically stable bound
state is possible for $m=1$ only.} \label{fig1}
\end{figure}

To verify the predictions of our effective-particle approach, we
perform a series of simulations of different $N$-soliton rings,
using the fast-Fourier-transform split-step numerical algorithm
and monitoring the conservation of the integrals of motion.
Alongside with non stationary behavior, such as breathing and
radiation emitting, we find the clusters dynamics in {\em
excellent agreement with our analysis}. According to Fig. 1, the
effective potential is always attractive for $m=0$, and thus the
ring of $N=5$ in-phase solitons should exhibit oscillations and,
possible, soliton fusion. Indeed, such a dynamics is observed in
Fig. 2(a). Although the oscillations of the ring are well
described by the effective potential $U(R)$, the ring dynamics is
more complicated. Another scenario of the mutual soliton
interaction corresponds to the repulsive potential [shown, e.g.,
in Fig. 1 for the case $m=2$ and $\theta=4\pi/5$]. In the
numerical simulations corresponding to this case, the ringlike
soliton array expands with the slowing down rotation, as is shown
in Fig. 2(d).

\begin{figure}
\vspace{-1 mm}\setlength{\epsfxsize}{7.5 cm} \centerline{
\epsfbox{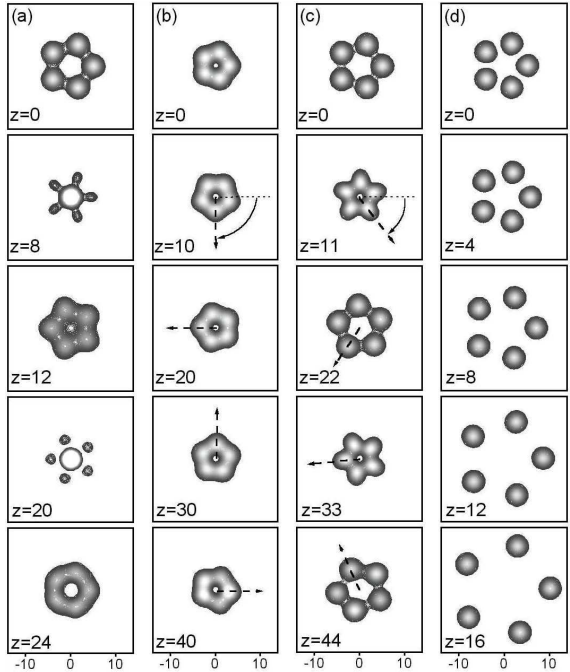}} \vspace{1.5 mm} \caption{Different regimes of
the interaction of $N=5$ solitons: (a) $m=0$, collapse and fusion
through oscillations; (b) $m=1$, a stationary bound state with
$R_{0}=3$; (c) $m=1$, an excited bound state with the oscillation
period $z_{\rm period}=22$; (d) $m=2$, the soliton repulsion. The
initial radius of the ring in the cases (a), (c), and (d) is
$R=5$.} \label{fig2}
\end{figure}

Evolution of the stationary bound state that corresponds to a minimum of the effective potential
$U(R)$ in Fig. 1 (for $m=1$) is shown in Fig. 2(b). Here the angular momentum is nonzero, and it
produces a repulsive centrifugal force that balances out the soliton attraction. The effective
potential predicts the stationary state at $R_{0}=3$ with a good accuracy, and the cluster does
not change its form while rotating during the propagation. To continue the analogy between the
soliton cluster and a rigid body, we calculate the cluster's {\em moment of inertia}, ${\cal I}$,
and its {\em angular velocity}, $\Omega$:
\begin{equation} \label{vel}\ {\cal I}=\int |E|^2 {\bf r}^2
d{\bf r},\;\;\Omega =M/{\cal I}.
\end{equation}
For the case shown in Fig. 2(b), the numerically obtained value of
the angular velocity is $\Omega_{num}\simeq\pi/20=0.157$, while
the formula (\ref{vel}) gives the value $\Omega = 0.154$. We also
perform the numerical simulations of the ``excited'' clusters, as
is shown in Fig. 2(c), and observed oscillations near the
equilibrium state. Such a {\em vibrational} state of the
``$N$-soliton molecule'' demonstrates the dynamical radial
stability of the bound state in agreement with the
effective-particle approximation.

Our analysis is valid for any $N$, and it allows us to classify
all possible scenarios of the soliton interaction in terms of the
phase jump $\theta$ between the neighboring solitons in the array.
Indeed, for $\theta=0$, the ring of $N$ solitons collapses through
several oscillations. If the main value of $\theta$ belongs to the
segment $0<\theta\leq\pi/2$, the interaction between solitons is
{\em attractive}, the value of the induced angular momentum is
finite, and there exists a rotating bound state of $N$ solitons.
However, if $\theta$ belongs to the segment $\pi/2<\theta\leq\pi$,
the soliton interaction is {\em repulsive} and the soliton ring
expands with or without ($\theta=\pi$) rotation, similar to the
necklace-type beams \cite{necklace}. For example, in the case
$N=3$ two stationary states are possible, $\theta=0$ and
$\theta=2\pi/3$, and there exist {\em no bound states}. For $N=4$
and $m=1$, the value of $\theta$ is $\pi/2$ and a cluster is
indeed possible, as is shown in Fig. 3.

\begin{figure} \setlength{\epsfxsize}{7.5 cm}
\centerline{ \epsfbox{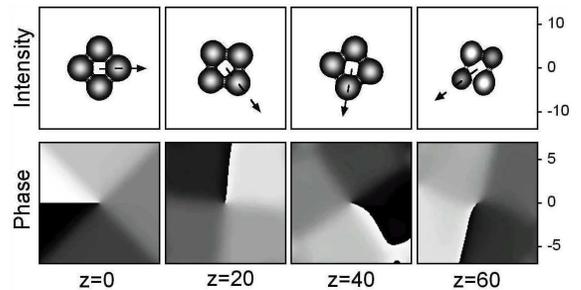}} \vspace{1.5 mm} \caption{Rotating
cluster of $N=4$ solitons. The parameters are $R_{0}=3.8$ and
$\Omega=0.042$. Phase images are scaled from $-\pi$ (black) to
$+\pi$ (white).} \label{fig3}
\end{figure}

Together with the intensity of the four-soliton cluster, in Fig. 3
we show the phase distribution for the distances up to $60$
diffraction lengths. The initial staircase-like phase in the ring
preserves its shape, and it is a nonlinear function of the polar
angle $\varphi$, similar to the phase of the necklace-ring vector
solitons with a fractional spin \cite{vecneck}. Note, that the
phase of such a state can be described as a {\em zeroth-order}
term in the expansion of the vortex phase near the $n$-th soliton
center:
\begin{equation}
\label{vort}\ m\varphi\simeq m\frac {2\pi n}{N} - m \frac
{y_{n}}{R^{2}} (x-x_{n})+m\frac {x_{n}}{R^{2}}(y-y_{n}) \ldots
\end{equation}
Calculating the minimum of the potential $U(R)$ that corresponds
to the soliton cluster, we find that for given $m$ and $N \gg 1$,
the stationary cluster approaches a vortex soliton of the charge
$m$. For example, the equilibrium radius $R_{0}$ for the clusters
with $m=1$ is $R_{0}=3.8$ for $N=4$, and for $N \geq 4$ it
approaches the corresponding vortex radius $R_{0}=3$. Furthermore,
for $m=2$, the soliton bound states are possible only if
$\theta=4\pi/N\leq\pi/2$. This gives the condition $N\geq 8$, and
for $N\geq9$ we find $R_{0}=5$ which is close to the radius of a
double-charged vortex. Thus, the ringlike soliton cluster can be
considered as a nontrivial ``discrete'' generalization of the
optical vortex soliton \cite{boris}. Clusters are generally
metastable, and they experience the symmetry-breaking instability.
However, they can propagate for many tenths of the diffraction
length, being also asymptotically stable in some types of
nonlinear media similar to the vortex solitons \cite{isaak}.

In Figs. 4(a,b,d) we show some examples of the excited rotating
clusters with different number of solitons $N$ and initial radius
$R=N$, while in Fig. 4(c) we present the stationary rotating
ringlike cluster of $N=8$ solitons.

\vspace{-1.5 mm}

\begin{figure} \setlength{\epsfxsize}{7.2 cm}
\centerline{ \epsfbox{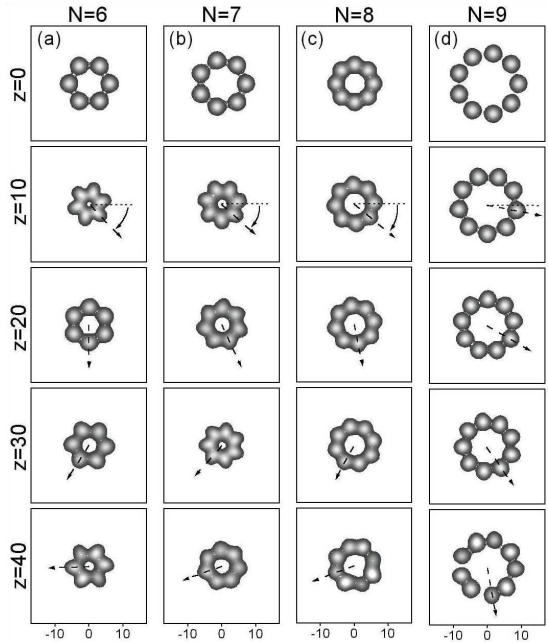}} \vspace{1.5 mm} \caption{Examples
of the rotating soliton clusters: (a) $N=6$ and $m=1$; (b) $N=7$
and $m=1$; (c) $N=8$ and $m=2$, here we show the exact bound state
with $R_{0}=5.5$ and the angular velocity $\Omega =0.092$; (d)
$N=9$ and $m=2$.} \label{fig4} \vspace{-1 mm}
\end{figure}

We should stress that the staircase-like phase distribution is a
distinctive feature of the soliton cluster. The evolution of the
soliton ring with the linear phases, i.e. those presented by the
first-order terms of Eq. (\ref{vort}), can be associated with
complex deformation of the vortex.

More interesting structures are found for the vector fields and
incoherent interaction of solitons. For example, three coherently
interacting solitons cannot form a bound state. However, adding a
single ``atom'' to the incoherently coupled additional component
$E_1$ [see Fig. 5(a)] leads to mutual trapping of all beams. Here,
the incoherent attraction balances out both the coherent
repulsion, as in the case of multipole vector solitons
\cite{anton}, and the centrifugal force induced by a net angular
momentum. This leads to the cluster rotation similar to the
two-lobe rotating ``propeller'' soliton \cite{propeller}. In Fig.
5(b), we show the five-lobe analogue of the necklace-ring vector
solitons recently discussed in Ref. \cite{vecneck}. Unlike the
necklace-ring solitons, the vector cluster rotates and undergoes
internal oscillations as it propagates.

\begin{figure}
\setlength{\epsfxsize}{7.0 cm} \centerline{ \epsfbox{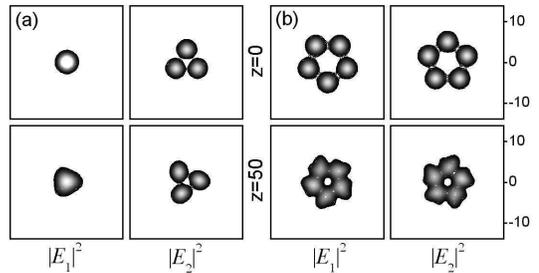}}
\vspace{1.0 mm} \caption{Examples of the vector soliton clusters.}
\label{fig5} \vspace{-2 mm}
\end{figure}

In conclusion, we have revealed a key physical mechanism for
stabilizing multi-soliton bound states in a bulk medium in the
form of rotating ringlike clusters. Such soliton clusters can be
considered as a nontrivial generalization of the important
concepts of the two-soliton spiraling, optical vortex solitons,
and necklace scalar beams, and they provide an example of the next
generation of multiple soliton-based systems operating entirely
with light. We believe that the basic ideas presented in this
Letter will be useful for other applications, such as the beam
dynamics in plasmas \cite{plasma} and the Skyrme model of a
classical field theory \cite{skyrme}, and they can be also
generalized to the inhomogeneous systems (such as the
Bose-Einstein condensates in a trap \cite{bec}).

The authors are indebted to M. Segev and M. Solja\^ci\'c for
encouraging comments and suggestions, and E.A. Ostrovskaya for a
help with numerical simulations and critical reading of this
manuscript.

\end{document}